\documentclass[aps, prd, preprint, groupedaddress,showpacs,18pt]{revtex4}
\usepackage{hyperref}
\usepackage{mathrsfs}
\usepackage{bbm}
\usepackage{amsfonts}
\usepackage{amsmath}
\usepackage{graphicx}
\usepackage{subfig}

\newcommand{\void}[1]{}

\renewcommand{\u}{\ensuremath{\mathfrak{u}}}

\def\be{\begin{equation}}
\def\ee{\end{equation}}
\def\bea{\begin{eqnarray}}
\def\eea{\end{eqnarray}}
\def\lmd{\lambda}
\def\la{\langle}
\def\ra{\rangle}
\def\bt{\beta}
\def\ga{\gamma}
\def\pt{\partial}
\newcommand{\ppt}[1]{{\partial\over \partial #1}}
\newcommand{\td}[1]{\tilde{#1}}
\newcommand{\dt}[1]{\dot{#1}~}

\newcommand{\bary}{\begin{array}}
\newcommand{\eary}{\end{array}}

\def\nb{\nonumber}
\begin{document}
\title{Spinor formalism for massive fields with half-integral spin}

\author{Gang~Chen$^{1}$, Konstantin G. Savvidy$^{1}$
}                     
%
\affiliation{
$^{1}$Department of Physics, Nanjing University\\
22 Hankou Road, Nanjing 210098, China
}
%

%
\begin{abstract}
In this article, we extend the 
spinor technique for calculating helicity amplitudes to general massive fields of half-integer spins. We find that the little group generators can be represented as first-order differential operators in the spinor formalism. We use the spinor forms of the generators to get the explicit form of the massive fields of any spin and any helicity. We also deal with the three-particle S-matrix by these spinor form generators, and find that we are able to extend the explicit form of the three-particle vertex obtained by Benincasa and Cachazo to the massive case. We present the explicit expressions for the amplitudes with external particles of the lowest helicities up to -3/2. Group theory, in the form of  raising operators of the little group, then dictates other amplitudes with higher helicity in the same spin multiplets. The formalism allows, in principle, to determine the electromagnetic form-factors of charged particles of arbitrary helicities, without additional assumptions about the underlying lagrangian. We find that restrictions which follow from gauge and Lorentz invariance are nearly as restrictive as in the massless case.
\end{abstract}
\date{\today}
\maketitle
\section{Introduction}
\label{intro}
The spinor formalism has been widely used in the calculation of amplitudes of four-dimensional gauge theories \cite{Parke:1986gb,Xu:1986xb,Berends:1987me,Dixon1,Dixon2,Dixon3,Dixon4,Wu:2004fba}. In a remarkable work \cite{Witten1}, Witten discovered that the tree-level amplitudes have a simple geometrical description when transformed into twistor space. This lead to a lot of new results on the properties of the amplitudes for massless particles. In particular, the S-matrices with more external particles can be deduced by the BCFW recursion relations \cite{Britto:2004nj,Britto:2004nc,Britto:2004ap,Britto:2005dg,Britto:2005fq,Britto:2005ha,Georgiou:2010mf}. Moreover,  the three-particle S-matrices of any spin can be uniquely determined \cite{Cachazo1}. Then the perturbative scattering amplitudes in Yang-Mills theory have a remarkably simple form. Many properties of the amplitudes are apparent in this form, even though they are not evident in the usual Feynman diagram computation.

For the massive case, the spinor form has also been used in simplifying algebra in Feynman graphs \cite{Dittmaier,Chalmers}. Also, the recursion relations for the gauge theory with massive particles have been developed in \cite{Badger1,FFF,Schwinn1,Ozeren,Schwinn2}. In this paper, we discuss the massive fields from another point of view.

The momentum of massive particles can be specified in terms of two Weyl spinors \cite{Witten1,Chalmers}. The ambiguity in the definition of the two spinors can be understood as the freedom to choose a basis of states for a massive particle, as this basis can be rotated under the little group. We also find the generators of the little group $SO(3)$ can be represented as first-order differential operators in these two spinors. 
By acting with the raising operator $J^+$ of $SO(3)$ on the lowest weight states, we can get the highest weight irreducible representations of the group. According to this property, we can get the spinor form of the free massive fields of any spin and helicity.

According to the spinor forms of the fields, we discuss the general property of the three-particle amplitudes. Here, we focus on the amplitude of two fermions and a massless gauge boson, where the fermions are of the same mass and arbitrary but equal half-integer spin. We get the general form of these amplitudes for the fermions of any helicity.  To achieve this, we first write down the first-order differential equations for the amplitudes, which are obtained by the linear property of the amplitudes with respect to each external field and the invariance of the external momenta under the little group for each particle. Then we construct the general solutions for the amplitudes when the massive external particles are of the lowest helicity structure. The amplitudes of other higher helicity structures can be obtained in the same way as we do for the free massive fields.

This paper is organized as follows. In Section \ref{MDfield}, we get the little group generators in the spinor form. Then we analyze in detail the amplitudes when external fermions are Dirac fields with spin ${1\over 2}$. In Section \ref{Spinthreehalf}, we use our method to the case when the external fermions are of spin $3\over 2$. In Section \ref{generalspin}, we extend to the case when fermions are of arbitrarily high spin. We find it is possible to get directly the spinor structure either for the external fields or the three-particle amplitudes.

\section{Massive  Dirac field}\label{MDfield}
For a massive field, the on-shell momentum $p^\mu$ are time-like with $p^2=m^2$, where $m$ is the mass of the particle.  Hence the momentum can be specified as $p_{a\dot{a}}=\lambda_a \td \lmd_{\dt a}+\bt_a \td\bt_{\dt a}$  in the bi-spinor form \cite{Kleiss1,Kleiss2,Novaes}, where $p_{a\dt a}=\sigma^\mu_{a\dt a}p_\mu$.  In these works, the arbitrariness in the decomposition is dealt with by fixing a phase, and one of the arbitrary momenta to some convenient value depending on the kinematics. 
In fact, the momentum is invariant when acting with the $U(2)$ group element $U$ on ${\lmd\choose\bt}$ and $U^\dagger$ on ${\td\lmd\choose \td\bt}$.   We explore and exploit the properties of the $SU(2)$ subgroup  and find a very simple way to write the spinor forms of free fields for any half-integral particles in Section \ref{generalspin}.  In the following, we fix the diagonal $U(1)$ phase by setting  the inner product of $\lmd$ and $\bt$ to be $\langle \lmd,\bt\rangle=[\td\lmd,\td\bt]=-m.$  Another motivation to fix the $U(1)$ phase is that this phase is only a parameter arbitrary without any physics content.  On the contrary, the $SU(2)$ is in fact the little group of the massive particles, which we discuss in detail in the following.  As noted in \cite{Chalmers}, this invariance is just the freedom to change basis of the solutions of free field equations, in the sense that the group invariance reflects both invariance of the momentum under the little group as well as the transformation properties of the solutions under the same group.  

The scattering amplitudes for particles of non-zero spin depend on not only the momentum but also the polarization vectors of the external particles. When the external particles are massive, we can also explicit the wave functions of them in $\lmd,\td\lmd,\bt,\td\bt$. Here we first consider the massive  Dirac fields. Such analysis can be extended to the massive on-shell vector fields of any spin.

The free field equation of  Dirac field is
\be
(i\ga^\mu\pt_\mu-m)\psi(x)=0.
\ee
For the positive frequency waves, $\psi(x)=u(p)e^{-ip\cdot x}$, the column vector $u(p)$ must obey
\be\label{diracEq}
(\ga^\mu p_\mu-m)u(p)=0,
\ee
where $p^2=m^2.$
We can divide the four component  Dirac spinor into two-component Weyl spinors as $u={\psi_L\choose\bar\chi_R}$. Then  equation (\ref{diracEq}) becomes
\bea
-m\psi_a +\sigma^\mu_{a\dt a}p_{\mu}\bar\chi^{\dt a}=0\nb\\
\bar\sigma_\mu^{a\dt a}p^{\mu}\psi_{a}-m\bar\chi^{\dt a}=0.
\eea
These have two independent solutions. We can choose the basis of the solutions as $u^{-}={\lmd_a\choose \td\bt^{\dt a}}$ and $u^{+}={-\bt\choose \td\lmd}$. 

In fact the group is directly related with the little group $SO(3)$ for massive particles. The relationship can be exhibited in the basis \footnote{Here and in the following, we usually omit the spinor index and the product of the spinors are taken as the tensor product unless otherwise specified.}
\bea\label{spinorbasis}
\textbf{E}_0&=&\lmd \td\lmd+\bt\td\bt \nb\\
\textbf{E}_3&=&\lmd \td\lmd-\bt\td\bt \nb\\
\textbf{E}_2&=&{1\over i}(\lmd \td\bt-\bt\td\lmd) \nb\\
\textbf{E}_1&=&\lmd \td\bt+\bt\td\lmd.
\eea
We can check directly that the four basis vectors are orthogonal to each other. Three of them are space-like vectors
$\textbf{E}_1,\textbf{E}_2,\textbf{E}_3$.
Another one is the time-like vector $\textbf{E}_0$, which is also proportional to the momentum $p_{a\dt a}$. By definition, the little group is the $SO(3)$ transformation of the three space-like vectors. The infinitesimal group elements generated by $L_3,L_2,L_1$ transform the vectors as
$(\textbf{E}_1\rightarrow \textbf{E}_1+\theta_3 \textbf{E}_2, \textbf{E}_2\rightarrow \textbf{E}_2-\theta_3 \textbf{E}_1, \textbf{E}_3\rightarrow \textbf{E}_3)$, $(\textbf{E}_3\rightarrow \textbf{E}_3+\theta_2 \textbf{E}_1, \textbf{E}_1\rightarrow \textbf{E}_1-\theta_2 \textbf{E}_3, \textbf{E}_2\rightarrow \textbf{E}_2)$, $(\textbf{E}_2\rightarrow \textbf{E}_2+\theta_1 \textbf{E}_3, \textbf{E}_3\rightarrow \textbf{E}_3-\theta_1 \textbf{E}_2, \textbf{E}_1\rightarrow \textbf{E}_1)$ respectively.

The two light-like vectors $\lmd_1\td\lmd_1$ and $\bt_1\td\bt_1$ also transforms correspondently, as shown explicitly in  Fig.1
\begin{figure}[htbp]
\begin{center}
\includegraphics[width=1.5in,angle=90]{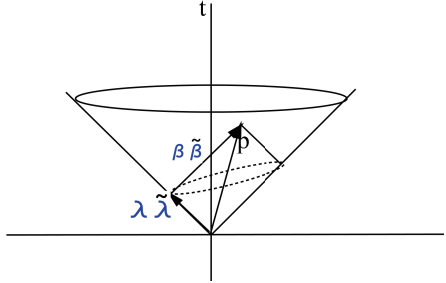}
\caption{{The dotted line indicates the orbit of the two light-like vectors under the $SU(2)$.  Only two generators transform the light-like vectors nontrivially.  The third one only applies a phase which is opposite for $\lmd$ and $\bt$.  }}
\label{lightcone}
\end{center}
\end{figure}

According to (\ref{spinorbasis}), we find the transformation of the space-like vectors under little group can be induced by the transformation of the spinors under the $SU(2)$, which also leave the momentum invariant. The corresponding infinitesimal transformations of the spinors are
\bea\label{spinortrans}
&&{\lmd\choose\bt}\rightarrow
{\lmd\choose\bt} \nb\\
&&-{i\over 2}\theta_3 \sigma_3 {\lmd\choose\bt}+{i\over 2}\theta_2 \sigma_2 {\lmd\choose\bt}-{i\over 2}\theta_1 \sigma_1 {\lmd\choose\bt}\nb\\
&&{\td\lmd\choose\td\bt}\rightarrow {\td\lmd\choose\td\bt}\nb\\ &&+{i\over 2}\theta_3 \bar\sigma_3 {\td\lmd\choose\td\bt}-{i\over 2}\theta_2 \bar\sigma_2 {\td\lmd\choose\td\bt}+{i\over 2}\theta_1 \bar\sigma_1{\td\lmd\choose\td\bt}.
\eea
Hence, the three generators of the little group $SO(3)$ can be represented as the first-order differential operators with respect to spinors
\bea
\mathcal{R}(J^1)&=&{-1\over 2} \left(\bt\ppt{\lmd}-\td\bt\ppt{\td\lmd}+\lmd\ppt{\bt}-\td\lmd\ppt{\td\bt} \right),\nb\\
\mathcal{R}(J^2)&=& {i\over 2} \left(\bt\ppt{\lmd}+\td\bt\ppt{\td\lmd}-\lmd\ppt{\bt}-\td\lmd\ppt{\td\bt} \right), \nb\\
\mathcal{R}(J^3)&=& {-1\over 2} \left(\lmd\ppt{\lmd}-\td\lmd\ppt{\td\lmd}-\bt\ppt{\bt}+\td\bt\ppt{\td\bt} \right).
\eea
Here, we make a convention on the summation of the spinor index in $\bt\ppt{\lmd}$ et al.  These spinor operators obey the usual commutation relations
\be
[\mathcal{R}(J^i), \mathcal{R}(J^j)]=i \epsilon_{ijk} \mathcal{R}(J^k).
\ee
Then according to (\ref{spinortrans}), the free Dirac field transforms as
\bea
&&{u^{-}\choose u^{+}} \rightarrow {u^{-}\choose u^{+}}\nb\\
&&-{i\theta_3\over 2}\sigma_3 {u^{-}\choose u^{+}}+{i\theta_2\over 2}\sigma_2 {u^{-}\choose u^{+}}-{i\theta_1\over 2}\sigma_1 {u^{-}\choose u^{+}}.
\eea
As expected, the two solutions of the Dirac equation are in the $\mathcal{D}^{1\over 2}$ representation of the little group. The two solutions can also be taken as the functions of the spinors. Hence the external free Dirac fields should obey the auxiliary conditions
\bea
\label{def:R}
\mathcal{R}(J^3) u_s &=&\mathcal{D}^{1\over 2}_{r s}(J^3)u_r,\nb\\
\mathcal{R}(J^2) u_s &=&\mathcal{D}^{1\over 2}_{rs}(J^2)u_r,\nb\\
\mathcal{R}(J^1)u_s &=&\mathcal{D}^{1\over 2}_{rs}(J^1)u_r.
\eea
Here, we denote $\mathcal{D}^{1\over 2}(J^i)$ as the two dimensional irreducible representation matrices for the generators of the little group. Respectively, $\mathcal{D}^{1\over 2}(J^1)={-1\over 2}\sigma_1$, $\mathcal{D}^{1\over 2}(J^2)={1\over 2}\sigma_2$, $\mathcal{D}^{1\over 2}(J^3)={-1\over 2}\sigma_3$.

It is convenient to use  another basis for this algebra, the $J^3$, $J^+=J^1+iJ^2$ and $J^-=J^1-iJ^2$. In the spinor form, the raising and lowering operators are $\mathcal{R}(J^+)= \left(\td\lmd\ppt{\td\bt}-\bt\ppt{\lmd} \right)$ and $\mathcal{R}(J^-)= \left(\td\bt\ppt{\td\lmd}-\lmd\ppt{\bt}\right)$ respectively. Actually, in this group of generators, we can exhibit the fields of any helicity as spinor forms. Usually, the field of the lowest or highest helicity can be obtained by the equations of motions and the little group representations. Then by acting with the raising or lowering operator on them, we can get the spinor forms for fields of any helicity. For example, in spin ${1\over 2}$ case, the highest helicity field can be obtained by acting a $\mathcal{R}(J^+)$ to the lowest helicity fields. That is $u^+= \mathcal{R}(J^+) u^-$. The procedure is similar with the one in constructing the highest weight irreducible representations of $SU(2)$ group. Furthermore, this procedure is also easy to generalize to the fields of any spin.

The scattering amplitudes $A(\lmd_1,\bt_1,\lmd_2,\bt_2,\lmd_3,\widetilde{\cdots})$ are functions of both the momenta $p_i$ and the wavefunctions $\u_i$ of each external particle. They are also linear with respect to the external fields. Then, due to the invariance of the momentum under little group, the amplitudes transform under the group for each particle in the same way as a free field. Here we focus on the amplitude $A({1\over 2}, {1\over 2}, 1)$, which is the interaction of two massive  Dirac fermions and one massless gauge boson. We find the amplitude satisfies the first-order differential relations
\bea\label{amponediff}
&&\mathcal{R}(J^a_1) A^{u_{r_1},u_{r_2},\pm}(\lmd_1,\bt_1,\lmd_2,\bt_2,\lmd_3,\widetilde{\cdots})\nb\\ &=&A^{\mathcal{D}^{{1\over2}}_{sr_1} u_{s}^1, u_{r_2}^2,\pm}(\lmd_1,\bt_1,\lmd_2,\bt_2,\lmd_3,\widetilde{\cdots}), \nb\\
&&\mathcal{R}(J^a_2) A^{u_{r_1},u_{r_2},\pm}(\lmd_1,\bt_1,\lmd_2,\bt_2,\lmd_3,\widetilde{\cdots})\nb\\ &=&A^{u_{r_1}^1,\mathcal{D}^{{1\over2}}_{sr_2} u_{s}^2, \pm}(\lmd_1,\bt_1,\lmd_2,\bt_2,\lmd_3,\widetilde{\cdots}), \nb\\
&&\mathcal{R}(J^3_3) A^{u_{r_1},u_{r_2},\pm}(\lmd_1,\bt_1,\lmd_2,\bt_2,\lmd_3,\widetilde{\cdots})\nb\\ &=&\pm A^{u_{r_1},u_{r_2},\pm}(\lmd_1,\bt_1,\lmd_2,\bt_2,\lmd_3,\widetilde{\cdots}).
\eea
Here, the up indexes of $A$ denote  the three external states respectively.  And we use $J^a_1$ and $J^a_2$ to denote the little group generators of the first and second particle respectively. For the photon, which is massless, only one generator $J^3_3$ of the little group acts non-trivially on the field. In total, there are seven first-order differential relations, three for each massive particle and one for the massless gauge boson.

We thus know the transformation properties of the amplitudes under the little group of each particle. Then, same as in the case of the free external fields, we do not need to obtain the general forms of the amplitudes for every helicity structure one by one. In fact, we can first choose that the states of  the massive particles  are all of the lowest helicity. Then, by acting with $\mathcal{R}(J^+)$ to the amplitude for these lowest helicity fields, we can obtain the amplitude for arbitrary helicity structure.

Now we take the two external  Dirac fields as $u_1^-, u_2^-$ and the external photon in negative helicity polarization. The three point  amplitude \\ $A\big(({1\over 2},-{1\over 2}), ({1\over 2},-{1\over 2}), (1,-1)\big)$ is non-zero only when the external fields are of complexified momenta. As in the massless case, we keep the on-shell conditions for the complexified momenta. In this section, we omit the spin index in the amplitudes while keeping only the helicity index in them.  We denote the momenta of the fermions and gauge boson as $p_1, p_2$ and $p_3$ respectively. Since the fermions are massive, as discussed above, we should at least use two independent spinors to characterize the momenta as $p^1_{a\dt a}=\lmd^1\td\lmd^1+\bt^1\td\bt^1$, $p^2_{a\dt a}=\lmd^2\td\lmd^2+\bt^2\td\bt^2$. The photon remains massless, $p^3_{a\dt a}=\lmd^3\td\lmd^3$.

Momentum conservation $(p_1+p_2+p_3)_{a\dt a}=0$ and the on-shell conditions $p_1^2=m^2, p_2^2=m^2, p_3^2=0$, imply that
\bea\label{conditions}
\langle\lmd_1,\lmd_2\rangle \, [\td\lmd_1,\td\lmd_2]+ \langle\lmd_1,\bt_2\rangle \, [\td\lmd_1,\td\bt_2]&&\nb\\
+ \langle\bt_1,\lmd_2\rangle \, [\td\bt_1,\td\lmd_2]+ \langle\bt_1,\bt_2\rangle \, [\td\bt_1,\td\bt_2] &=& -2m^2, \nb\\
\langle\lmd_1,\lmd_3\rangle \, [\td\lmd_1,\td\lmd_3]+ \langle\bt_1,\lmd_3\rangle \, [\td\bt_1,\td\lmd_3]&=&0, \nb\\
\langle\lmd_2,\lmd_3\rangle \, [\td\lmd_2,\td\lmd_3]+ \langle\bt_2,\lmd_3\rangle \, [\td\bt_2,\td\lmd_3]&=&0.
\eea

By Lorentz invariance, the scattering amplitudes can only be the functions of the eight invariants $\langle\lmd_1,\lmd_2\rangle$,$\langle\lmd_3,\lmd_2\rangle$, $\langle\lmd_1,\lmd_3\rangle$, $\langle\bt_1,\lmd_2\rangle$, $\langle\lmd_1,\bt_2\rangle$, $\langle\bt_1,\bt_2\rangle$, $\langle\bt_1,\lmd_3\rangle$, $\langle\bt_2,\lmd_3\rangle,$ and the corresponding right-handed parts. However, only ten of the total of sixteen variables are independent. Without loss of generality, we assume $\langle\lmd_1,\lmd_2\rangle\neq 0$ and $[\td\lmd_1,\td\lmd_2]\neq 0$. Then the first five of left handed and right handed variables are independent. This can be seen by defining the identity "operators" as $\mathbb{I}_L={,\lmd_2\rangle \, \langle\lmd_1,-,\lmd_1\rangle \, \langle\lmd_2,\over \langle\lmd_1,\lmd_2\rangle}$ and $\mathbb{I}_R={,\td\lmd_2][\td\lmd_1,- ,\td\lmd_1][\td\lmd_2,\over [\td\lmd_1,\td\lmd_2]}$, which  act on the left and right handed spinors respectively. By inserting $\mathbb{I}_L$ into the inner product $\langle\bt_1,\bt_2\rangle, \langle\bt_1,\lmd_3\rangle, \langle\bt_2,\lmd_3\rangle$, we can get
express these three invariants in terms of the other five:
\bea
\langle\bt_1,\bt_2\rangle&=&{m^2+\langle\bt_1,\lmd_2\rangle \,  \langle\lmd_1,\bt_2\rangle \, \over \langle\lmd_1,\lmd_2\rangle},\nb\\ \langle\bt_1,\lmd_3\rangle&=&{-m\langle\lmd_2,\lmd_3 \rangle +\langle\bt_1,\lmd_2\rangle \,  \langle\lmd_1,\lmd_3\rangle\over \langle\lmd_1,\lmd_2\rangle},\nb\\
 \langle\bt_2,\lmd_3\rangle&=&{\langle\lmd_1,\bt_2\rangle \,  \langle\lmd_2,\lmd_3\rangle +m \, \langle\lmd_1,\lmd_3\rangle\over \langle\lmd_1,\lmd_2\rangle}.
\eea
Similarly, according to $\mathbb{I}_R$, we can get
\bea
~[\td\bt_1,\td\bt_2]&=&{m^2+[\td\bt_1,\td\lmd_2] [\td\lmd_1,\td\bt_2]\over [\td\lmd_1,\td\lmd_2]},\nb\\
~[\td\bt_1,\td\lmd_3]&=&{-m[\td\lmd_2,\td\lmd_3 ] +[\td\bt_1,\td\lmd_2] [\td\lmd_1,\td\lmd_3]\over [\td\lmd_1,\td\lmd_2]},\nb\\
~ [\td\bt_2,\td\lmd_3]&=&{[\td\lmd_1,\td\bt_2] [\td\lmd_2,\td\lmd_3] +m [\td\lmd_1,\td\lmd_3]\over [\td\lmd_1,\td\lmd_2]}.
\eea

In the spinor form, the photon with negative helicity is $\epsilon_{a\dt a}={\lmd_a \td \mu_{\dt a} \over [\td\lmd, \td\mu]}$ \cite{Witten1}.

Hence in principle there are enough constraints to solve the differential equations. We now construct the possible solutions directly from the linearity property of the amplitudes with respect to the external fields.
Momenta are invariant under the action of the differential operators (\ref{def:R}). Then if an amplitude is a solution of (\ref{amponediff}), it will still satisfy the equations when inserted in any momentum function. Meanwhile, the amplitude, which is composed of the tensor product of each external field,  is a solution of (\ref{amponediff}). Then we construct the solutions in terms of some undetermined functions of momenta
and the spinor forms of the fields such that the expressions for the amplitudes are Lorentz invariant.

 When the helicity of the external gluon is minus, the linear property will lead to $\lmd_1$ or $\td\bt_1$ and  $\lmd_2$ or $\td\bt_2$ and ${\lmd_3\td\mu \over [\td \mu, \td \lmd_3]}$ be of order one in the amplitude.  To form the Lorentz invariant terms,  they  should be constructed by spinor product directly or with the spinor forms of the external momentums.   If the spinors of in the external states are of the same charity, the external momentums should be of even orders to form a Lorentz invariant factor, i.e. 
 \bea
 &&\langle \lmd_1| \, F_1^e(p_1,p_2)|\lmd_2\rangle \nb\\
 &= &\lmd_1^{a} \, p^1_{a_1\dot{a}_1}p_1^{\dot{a}_1 a_2}\cdots p^{2}_{a_{2n-2}\dot{a}_{2n-1}}p_{2}^{\dot{a}_{2n-1} a_{2n}} \, \lmd^2_{a_{2n}}.\nb\\ 
 \eea
 Here and following the external momentum $p_1, p_2$ are arranged such that $p_1$ are all in the left hand of $p_2$. This is possible since the order of $p^1\circ p^2\equiv p^1_{a_1\dot{a}_1}p_2^{\dot{a}_1 a_2}$ can be changed  up to an $\delta_{a_1}^{a_2}$ due to  
 $\sigma^\mu \bar\sigma^\nu+\sigma^\nu \bar \sigma^\mu=2 \delta_{a_1}^{a_2}\eta^{\mu\nu}$. 
 
 Otherwise, the external momentums should be of odd orders, i.e. 
 \bea
 &&\langle \lmd_1| \, F_1^o(p_1,p_2)|\td\mu] \nb\\
 &= &\lmd_1^{a} \, p^1_{a_1\dot{a}_1}p_1^{\dot{a}_1 a_2}\cdots p^{2}_{a_{2n-2}\dot{a}_{2n-1}}\, \td \mu^2_{\dot{a}_{2n-1}}.\nb\\ 
 \eea 
The factors include $\td\bt_{1,2}$ can be  also write in the general factors with only $\lmd_{1,2}$. This is due to \\
$p^{1,2}_{a\dt a}|\td\bt_{1,2}^{\dt a}\rangle=|\lmd^{1,2}_a\rangle$. Hence,  the general form of the amplitude should be 
\bea\label{genAmp}
A~=&
 \big(\vphantom{\sum}\big.&
 c_1\,{\langle \lmd_1|\,F_1^e(p_1,p_2)|\lmd_2\rangle ~ [\td \mu|\,F_2^o(p_2)|\lmd_3\rangle }  \nb\\
&+&c_2\,{\langle \lmd_1|\,F_3^o(p_1,p_2)|\,\td\mu\,]~\langle\lmd_2, \lmd_3\rangle }\nb\\
&+&c_3\,{\langle \lmd_2|\,F_4^o(p_1,p_2)|\,\td\mu\,]~\langle\lmd_1, \lmd_3\rangle }\big. \vphantom{\sum}\big) ~ \frac{1}{ [\td \mu, \td \lmd_3]}\nb\\
\eea

Since in gauge theory, the amplitudes are on-shell gauge invariance.  We can find that the   first term in (\ref{genAmp}) is gauge invariant by itself.  Actually, under $\td\mu \rightarrow \td\mu+\td\lmd_3$, the factor  $[\td \mu|\,F_2^o(p_2)|\lmd_3\rangle$ is invariant due to the momentum conservation and on-shell condition of the external lines.  However  the last two terms in (\ref{genAmp}) are not gauge invariant 
in general. It is possible to rewrite the general forms $F_{3}^o(p_1,p_2)$ and  $F_{4}^o(p_1,p_2)$ into  $F_{3}^o(p_1+p_2, p_1)$ and $F_{4}^o(p_1+p_2, p_2)$ respectively.  Furthermore, since the we can change the orders of $p_1+p_2, p_{1}$ or $p_1+p_2, p_{2}$ freely,  $F_{3}^o(p_1+p_2, p_1)$ can be classified  as  $F_{3}^e(p_1) \circ F_{3}^o(p_1+p_2), F_{3}^o(p_1)\circ F_{3}^e(p_1+p_2)$.  And since the external momentum of gluon is light-like, $F_{3}(p_1+p_2)$ can at most of order one for the momentum.  As a result, the $F_{3}^o(p_1,p_2)$ can be simplified to  $F_{3}^e(p_1) \circ (p_1+p_2), F_{3}^o(p_1)$. Similarly, $F_{4}^o(p_1+p_2, p_2)$ can be simplified to $F_{4}^e(p_2) \circ (p_1+p_2), F_{4}^o(p_2)$. 

We find for the terms with factors 
\bea
 {\langle \lmd_1|\,F_{3}^e(p_1) \circ (p_1+p_2)|\,\td\mu\,]~\langle\lmd_2, \lmd_3\rangle },\nb\\
 {\langle \lmd_1|\,F_{4}^e(p_2) \circ (p_1+p_2)|\,\td\mu\,]~\langle\lmd_2, \lmd_3\rangle },\nb
\eea
they are unchanged under $\td\mu \rightarrow \td\mu+\td\lmd_3$ and hence on-shell gauge invariant. 
The terms with $F_{3}^o(p_1),F_{4}^o(p_2)$ are not gauge invariant directly. But a combination of them can still be gauge invariant 
\bea
\langle \lmd_1|\,F_5^o(p_1)|\,\td\mu\,]~\langle\lmd_2,\lmd_3 \rangle +\,\langle \lmd_2|\,F_{5}^o(p_2)|\,\td\mu\,]\,\langle\lmd_1,\lmd_3 \rangle
\vphantom{\sum}.\nb
\eea

Hence under the on-shell gauge invariant gauge constraints,  the allowed amplitudes are of forms
\bea\label{gengiAmp}
&&A~=
 \big(\vphantom{\sum}\big.
 c_1\,{\langle \lmd_1|\,F_1^e(p_1,p_2)|\lmd_2\rangle ~ [\td \mu|\,F_2^o(p_2)|\lmd_3\rangle }  \nb\\
&+&c_2\,{\langle \lmd_1|\,F_{3}^e(p_1) \circ (p_1+p_2)|\,\td\mu\,]~\langle\lmd_2, \lmd_3\rangle }\nb\\
&+&c_3\,{\langle \lmd_2|\,F_{4}^e(p_2) \circ (p_1+p_2)|\,\td\mu\,]~\langle\lmd_1, \lmd_3\rangle }\nb\\
&+&c_4\,(\langle \lmd_1|\,F_5^o(p_1)|\,\td\mu\,]~\langle\lmd_2,\lmd_3 \rangle \nb\\
&&~~~+\,\langle \lmd_2|\,F_{5}^o(p_2)|\,\td\mu\,]\,\langle\lmd_1,\lmd_3 \rangle)
\big. \vphantom{\sum}\big) \nb\\&\times&~ \frac{1}{ [\td \mu, \td \lmd_3]}.
\eea
We can further simplify the amplitude expression.  As discussed above, all the four terms in (\ref{gengiAmp}) are on-shell gauge invariant independently.  Moreover, in each term the order of $|\td\mu]$ in the numerator and  denominator is same. Hence we can choose a convenient value of $|\td\mu]$ other than $\td\lmd_3$ for each term separately. For convenience we set $|\td\mu]=\td\bt_2$.  Then (\ref{gengiAmp} becomes 
\bea\label{gengiAmp1}
&&A~=
 \big(\vphantom{\sum}\big.
 c_1\,{\langle \lmd_1|\,(1+p_1\circ p_2)|\lmd_2\rangle ~ [\td \bt_2|\,p_2|\lmd_3\rangle }  \nb\\
&+&c_2\,{\langle \lmd_1|\, (p_1+p_2)|\,\td\bt_2\,]~\langle\lmd_2, \lmd_3\rangle }\nb\\
&+&c_3\,{\langle \lmd_2|\, (p_1+p_2)|\,\td\bt_2\,]~\langle\lmd_1, \lmd_3\rangle }\nb\\
&+&c_4\,(\langle \lmd_1|\,p_1|\,\td\bt_2\,]~\langle\lmd_2,\lmd_3 \rangle +\,\langle \lmd_2|\,p_2|\,\td\bt_2\,]\,\langle\lmd_1,\lmd_3 \rangle)
\big. \vphantom{\sum}\big) \nb\\&\times&~ \frac{1}{ [\td \bt_2, \td \lmd_3]}.
\eea
Here we removing all the even orders of the momenta by acting it on the corresponding spinors. 
Then using the spinor forms of the momenta, we can calculate the general form of the amplitude directly
\bea\label{Amphalf1}
A({-1\over 2}, {-1\over 2}, -1)&=&{P^1(\langle\lmd_1,\lmd_2\rangle,[\td\bt_1,\td\bt_2]) ~
\langle\lmd_2,\lmd_3\rangle \over [\td \bt_2, \td \lmd_3]}\nb\\
&+&Q^0 \langle\lmd_1,\lmd_3\rangle \,  \langle\lmd_2,\lmd_3\rangle.
\eea

Here $Q^0$ only depends on the mass and $P^1$ is a first-order polynomial function:
$$a \, \langle\lmd_1,\lmd_2\rangle+b \, [\td\bt_1,\td\bt_2],$$ where $a, b$ are constants.  However, the terms in (\ref{Amphalf1}) are not independent with each other. In fact, using momentum conservation, we find
\be\label{Constraint}
\la\lmd_3,\lmd_2\ra \, [\td\lmd_3,\td\bt_1]=m \, ( \la\lmd_1,\lmd_2\ra +[\td\bt_1,\td\bt_2]).
\ee
Hence the amplitudes with independent solutions can be written as
\bea\label{Amphalf2}
A({-1\over 2}, {-1\over 2}, -1)&=&F_1 \, { [\td\bt_1,\td\bt_2]\langle\lmd_2,\lmd_3\rangle \over [\td \bt_2, \td \lmd_3]}\nb\\
&+&{F_2\over 2m}\langle\lmd_1,\lmd_3\rangle \,  \langle\lmd_2,\lmd_3\rangle.
\eea
In QED, it is easy to check that the term with coefficient $F_1$ comes from the minimal interaction
$\bar v_2({-1\over 2})\gamma_\mu u_1({-1\over 2})\,\epsilon_\mu$.
The term with the coefficient $F_2$ comes from the anomalous magnetic-moment interaction
$\bar v_2({-1\over 2}){i\sigma^{\mu\nu}p^3_\nu \over 2m} u_1({-1\over 2})$.

As discussed above, the amplitudes of other helicities are easy to obtain, according to (\ref{amponediff})
\bea\label{amppmmhalf}
A({1\over 2}, {-1\over 2}, -1)&=&F_1 \, {[\td\lmd_1,\td\bt_2] \, \langle\lmd_2, \lmd_3\rangle\over [\td\bt_2,\td\lmd_3]}\nb\\
&-&{F_2\over 2m}\,\langle\bt_1,\lmd_3\rangle \,  \langle\lmd_2,\lmd_3\rangle, \nb \\
A({-1\over 2}, {1\over 2}, -1)&=&F_1 \, {[\td\bt_1,\td\lmd_2] \, \langle\lmd_2, \lmd_3\rangle\over [\td\bt_2,\td\lmd_3]}\nb\\
&-&{F_2\over 2m} \,\langle\lmd_1,\lmd_3\rangle \,  \langle\bt_2,\lmd_3\rangle, \nb \\
A({1\over 2}, {1\over 2}, -1)&=&F_1 \, {[\td\lmd_1,\td\lmd_2] \, \langle\lmd_2, \lmd_3\rangle\over [\td\bt_2,\td\lmd_3]}\nb\\
&+&{F_2\over 2m}\,\langle\bt_1,\lmd_3\rangle \,  \langle\bt_2,\lmd_3\rangle.
\eea

If the anomalous magnetic interaction vanishes, $F_2=0$, then in the massless limit $\bt_i \rightarrow 0$, the fermion helicity is conserved and the amplitudes tend to
\bea
A_{m\rightarrow 0}(-{1\over 2}, -{1\over 2}, -1)&=&
0,\nb\\
A_{m\rightarrow0}({1\over 2}, {-1\over 2}, -1)&=&b_0\, {\langle\lmd_2, \lmd_3\rangle^2 \over \langle\lmd_1,\lmd_2\rangle},\nb\\
A_{m\rightarrow0}({-1\over 2}, {1\over 2}, -1)&=&b_0\, {\langle\lmd_1, \lmd_3\rangle^2 \over \langle\lmd_1,\lmd_2\rangle},\nb\\
A_{m\rightarrow0}({1\over 2}, {1\over 2}, -1)&=&0.
\eea
The results can be verified by the direct calculation using Feynman rules of QED.

\section{Massive spin ${3\over 2}$ fields in spinor form}\label{Spinthreehalf}
In this section we discuss the properties of the massive spin ${3\over 2}$ particle in the spinor form.  The paper of Novaes and Spehler \cite{Novaes} treats a specific form for the interaction of a spin 3/2 particle, namely the minimal interaction, whereas we are interested in obtaining the amplitude with the most general form factors. We first get the wave functions for the massive fields of spin ${3\over 2}$ irreducible representation of the little group. Then we analyze the general form of the three-point amplitudes with two massive spin ${3\over 2}$ fields and one massless gauge boson, where the mass of the two fermions are equal. The spin ${3\over 2}$ fields can be obtained from the vector-spinor $\psi_\mu$ \cite{Fierz,Rarita}
 \bea
 \psi_\mu&\simeq&({1\over 2},{1\over 2})\otimes\left[({1\over 2},0)\oplus (0,{1\over 2})\right]\nb\\
 &\simeq&(1,{1\over 2})\oplus({1\over 2},1)\oplus(0,{1\over 2})\oplus({1\over 2},0).
 \eea
In field theory, particles are classified by the irreducible representation of the little group. In massive case, the little group is $SO(3)$. Hence, a vector field in $({1\over 2},{1\over 2})$ contains two irreducible representations $\textbf{1}$ and $\textbf{0}$ of $SO(3)$. The spinor form of the vector field are defined in (\ref{spinorbasis}). Fields in $(0,{1\over 2})$ or $(0,{1\over 2})$ are an irreducible representation ${\textbf{1}\over \textbf{2}}$ with respect to the little group. Hence a spin ${3\over 2}$ particle should lie in the $\textbf{1}$ representation for the vector part and in the symmetric part of the tensor product between the vector and spinor. According to this group representation analysis, we can obtain the spinor form of the spin ${3\over 2}$ field directly.
For the  Dirac fields, the spinor forms of the wave functions are
\be
u^{-}={\lmd\choose \td\bt},~~u^{+}={-\bt\choose \td\lmd}, ~~v^-={\lmd\choose -\td\bt}, ~~v^+={\bt\choose \td\lmd}.
\ee
Here $u$ is to denote the particles and $v$ is to denote the anti-particles. Then by symmetrizing the tensor product of the spin $\textbf{1}$ vectors in (\ref{spinorbasis}) and the  Dirac fields, we get the wave functions of the spin ${3\over 2}$ fields as follows
\bea\label{uspinor}
u_{-3\over 2}&=&-\lmd \, \td\bt \, (\lmd \oplus\td\bt)\nb\\
u_{-1\over 2}&=&(-\lmd \, \td\lmd +\bt\, \td\bt) \, (\lmd \oplus\td\bt) - \lmd\, \td\bt\, (-\bt\oplus \td\lmd)\nb\\
u_{1\over 2}&=&-(-\lmd \, \td\lmd +\bt\, \td\bt) \, (-\bt \oplus \td\lmd) - \bt\, \td\lmd\, (\lmd\oplus \td\bt)\nb\\
u_{3\over 2}&=&\bt \, \td\lmd (-\bt \oplus \td\lmd)
\eea
For anti-particles, the spinor forms are
\bea\label{vspinor}
v_{-3\over 2}&=&-\lmd \, \td\bt \, (\lmd \oplus -\td\bt)\nb\\
v_{-1\over 2}&=&  (-\lmd \, \td\lmd +\bt\td\bt) \, (\lmd \oplus -\td\bt) + \lmd\, \td\bt\, (\bt\oplus \td\lmd)  \nb\\
v_{1\over 2}&=&   (-\lmd \, \td\lmd +\bt\td\bt) \, (\bt \oplus \td\lmd)  - \bt\, \td\lmd\, (\lmd\oplus -\td\bt) \nb\\
v_{3\over 2}&=&-\bt \, \td\lmd \, (\bt \oplus \td\lmd).
\eea
Here the states are classified by the eigenvalues of the $J_3$ spin which is also the spin components in the $E_3$ direction. 
We can check directly the conditions imposed by Pauli and Fierz in \cite{Fierz}. In fact, the use of the transverse vector parts of spin $\textbf{1}$ leads to the condition $p_\mu \, \psi^\mu=0$. Also, taking a symmetric tensor product reproduces the condition $\gamma_\mu \, \psi^\mu=0$.  

The spinor forms were also presented before in (\cite{Novaes,Feng}). Here we can also provide a very simple way to get them by action of the little group in the spinor form.  Such method can also be generalized to higher spin fields directly.  First, the lowest helicity field can only be $\lmd\td\bt (\lmd\oplus\td\bt)$ for positive-frequency particle. Then acting with $\mathcal{R}(J_1^+)$ on it several times, we can get the fields of higher helicity. The result is precisely (\ref{uspinor}). For the anti-particle, the procedure is similar.

Now we can analyse the amplitudes $A({3\over 2},{3\over 2}, 1)$ for two spin $3\over 2$  Dirac  fields and one gauge boson without any information on the interaction terms in Lagrangian. The external fields of spin $3\over 2$ are in the  representation $\mathcal{D}^{3\over 2}$ for the little group. Then the fields should satisfy the equations
\bea
\mathcal{R}(J^3) \, u_r&=&\mathcal{D}^{3\over 2}_{sr}(J^3)u_s\nb\\
\mathcal{R}(J^2) \, u_r&=&\mathcal{D}^{3\over 2}_{sr}(J^2)u_s\nb\\
\mathcal{R}(J^1) \, u_r&=&\mathcal{D}^{3\over 2}_{sr}(J^1)u_s,
\eea
Since the amplitudes are linear in the external fields, they should satisfy the similar equations
\bea\label{diffAmp}
\mathcal{R}(J^a_1) A^{u_{r_1},u_{r_2},\pm}&=&A^{\mathcal{D}^{{3\over2}}_{sr_1} u_{s}^1, u_{r_2}^2,\pm} \nb\\
\mathcal{R}(J^a_2) A^{u_{r_1},u_{r_2},\pm} &=&A^{u_{r_1}^1,\mathcal{D}^{{3\over2}}_{sr_2} u_{s}^2, \pm}, \nb\\
\mathcal{R}(J^3_3) A^{u_{r_1},u_{r_2},\pm} &=&\pm A^{u_{r_1},u_{r_2},\pm}.
\eea

As the case in Section \ref{MDfield}, we can construct the solutions of the different equations directly. Here we also set the two fermions to be of equal mass $m$ and the gauge boson is massless.

First we choose the helicity of each particle as $-{3\over 2}, -{3\over 2}, -1$. The general solution for the amplitude $A\big(({3\over 2},-{3\over 2}),({3\over 2},-{3\over 2}), (1,-1)\big)=A_1+A_2$, where 
\bea
&&A_1=\Big(\left(\right.v_1 \, [\td\bt_1| \, F_1^e(p_1,p_2)|\,\td\mu\,]\,\langle\lmd_1 | \, F_2^e(p_1,p_2)|\lmd_2\rangle \, \nb\\
&&\times\langle\lmd_3,\lmd_2\rangle \,  \langle\lmd_1 | \, F_3^o(p_1,p_2)|\td\bt_2] \nb\\
&&+v_{1'} \, [\td\bt_2| \, F_{1'}^e(p_1,p_2)|\,\td\mu\,]\,\langle\lmd_1| \, F_{2'}^e(p_1,p_2)|\lmd_2\rangle \, \nb\\
&&\times \langle\lmd_3,\lmd_1\rangle \,  \langle\lmd_2 | \, F_{3'}^o(p_1,p_2)|\td\bt_1]\,\left.\right)\nb\\
&+&\left(\right.v_2 \, [\td\bt_1| \, F_4^e(p_1,p_2)|\,\td\mu\,]\,\langle\lmd_1| \, F_5^e(p_1,p_2)| \lmd_2\rangle \, \nb\\
&&\times\langle\lmd_1,\lmd_2\rangle \,  \langle\lmd_3 | \, F_6^o(p_2)|\td\bt_2] \nb\\
&&-v_{2'} \, [\td\bt_2| \, F_{4'}^e(p_1,p_2)|\,\td\mu\,]\,\langle\lmd_1| \, F_{5'}^e(p_1,p_2)|\lmd_2\rangle \,  \nb\\
&&\times\langle\lmd_1,\lmd_2\rangle \,  \langle\lmd_3 | \, F_{6'}^o(p_1)|\td\bt_1]\,\left.\right)\nb\\
&+&\left(\right.v_3[\td\bt_1| \, F_7^e(p_1,p_2)|\,\td\mu\,]\,\langle\lmd_1| \, F_8^e(p_1,p_2)|\lmd_2\rangle \,  \nb\\
&&\times\langle\lmd_1,\lmd_3\rangle \,  \langle\lmd_2 | \, F_9^o(p_1)|\td\bt_2] \nb\\
&&-v_{3'}[\td\bt_2| \, F_{7'}^e(p_1,p_2)|\,\td\mu\,]\,\langle\lmd_1| \, F_{8'}^e(p_1,p_2)|\lmd_2\rangle \,  \nb\\
&&\times\langle\lmd_2,\lmd_3\rangle \,  \langle\lmd_1 | \, F_{9'}^o(p_2)|\td\bt_1]\,\left.\right)\nb\\
&+&\left(\right.v_4[\td\bt_1| \, F_{10}^e(p_1,p_2)|\td\bt_2]\,\langle\lmd_1| \, F_{11}^e(p_1,p_2)|\lmd_2\rangle \,  \nb\\
&&\times\langle\lmd_3,\lmd_2\rangle \,  \langle\lmd_1 | \, F_{12}^o(p_1)|\,\td\mu\,]\nb\\
&&-v_{4'}[\td\bt_2| \, F_{10'}^e(p_1,p_2)|\,\td\mu\,]\,\langle\lmd_1| \, F_{11'}^e(p_1,p_2)|\lmd_2\rangle \, \nb\\
&&\times \langle\lmd_3,\lmd_1\rangle \,  \langle\lmd_2 | \, F_{12'}^o(p_2)|\,\td\mu\,]\,\left.\right)\nb\\
&+&v_5[\td\bt_1| \, F_{13}^e(p_1,p_2)|\td\bt_2]\,\langle\lmd_1| \, F_{14}^e(p_1,p_2)|\lmd_2\rangle \,  \nb\\
&&\times \langle\lmd_3,\lmd_2\rangle \,  \langle\lmd_1 | \, F_{15}^o(p_1+p_2)|\,\td\mu\,] \nb\\
&+&v_6[\td\bt_2| \, F_{16}^e(p_1,p_2)|\td\bt_2]\,\langle\lmd_1| \, F_{17}^e(p_1,p_2)|\lmd_2\rangle \,  \nb\\
&&\times \langle\lmd_3,\lmd_1\rangle \,  \langle\lmd_2 | \, F_{18}^o(p_1+p_2)|\,\td\mu\,]\nb\\
&+&v_7[\td\bt_1| \, F_{19}^e(p_1,p_2)|\td\bt_2]\,\langle\lmd_1| \, F_{20}^e(p_1,p_2)|\lmd_2\rangle \, \nb\\
&&\times \langle\lmd_1| \, F_{21}^e(p_1,p_2)|\lmd_2\rangle \,  \langle\lmd_3 | \, F_{22}^o(p_1,p_2)|\,\td\mu\,]\Big)/[\td\mu,\td\lmd_3]\nb\\
\eea
and 
\bea
&&A_2=\la\lmd_1|F_{23}^o(p_2)|\bt_1] \la\lmd_2|F_{24}^o(p_1)|\bt_2]\nb\\
&&\times \big(\vphantom{\sum}\big.
 c_1\,{\langle \lmd_1|\,F_{25}^e(p_1,p_2)|\lmd_2\rangle ~ [\td \mu|\,F_{26}^o(p_1,p_2)|\lmd_3\rangle }  \nb\\
&&+c_2\,{\langle \lmd_1|\,F_{27}^o(p_1+p_2)|\,\td\mu\,]~\langle\lmd_2, \lmd_3\rangle }\nb\\
&&+c_3\,{\langle \lmd_2|\,F_{28}^o(p_1+p_2)|\,\td\mu\,]~\langle\lmd_1, \lmd_3\rangle }\nb\\
&&+c_4\,\langle \lmd_1|\,F_{29}^o(p_1)|\,\mu\,]~\langle\lmd_2,\lmd_3 \rangle \nb\\
&&\times +c_{4'}\,\langle \lmd_2|\,F_{29'}^o(p_2)|\,\mu\,]\,\langle\lmd_1,\lmd_3 \rangle
\big. \vphantom{\sum}\big) ~ \frac{1}{ [\td \mu, \td \lmd_3]}
\eea
Here, $F^{e/o}_{i}$ and $F^{e/o}_{i'}$ are
products of arbitrary even (odd) number momenta in its argument.
arbitrary functions of the momenta of even or odd order. Note also, that $v_i$ and $v_{i'}$ are constants independent of momenta. We can check that each term in the amplitude is a solution of the equations. Meanwhile, when $F_{i}=F_{i'}$ and $v_i=v_{i'}, c_i=c_{i'}$, they are all independent of the $\td\mu$ which indicates the on-shell gauge invariance of the theory. Hence we can choose $\td\mu$ to be arbitrary spinor not proportional to $\td\lmd_3$. Without loss of generality, we set $\td\mu=\td\bt_2$. Then according to momentum conservation and the on-shell conditions (\ref{conditions}), the amplitude can be simplified to
\bea\label{ampthreehalf}
&&A\big(({3\over 2},-{3\over 2}),({3\over 2},-{3\over 2}), (1,-1)\big)=\nb\\
&&{P^3\left([\td\bt_1,\td\bt_2],\langle\lmd_1,\lmd_2\rangle\right) \langle\lmd_2,\lmd_3\rangle \over [\td\bt_2, \td\lmd_3]}\nb\\
&+&\left(\la\lmd_1,\lmd_2\ra \, [\td\lmd_2,\td\bt_1]+\la\lmd_1,\bt_2\ra \, [\td\bt_2,\td\bt_1]\right)\nb\\
&&\times\left(\la\lmd_2,\lmd_1\ra \, [\td\lmd_1,\td\bt_2]+\la\lmd_2,\bt_1\ra \, [\td\bt_1,\td\bt_2]\right)\nb\\
&&\times P^1\left([\td\bt_1,\td\bt_2], \langle\lmd_1,\lmd_2\rangle\right){\langle\lmd_2,\lmd_3\rangle \over [\td\bt_2, \td\lmd_3]},
\eea
where $P^3$ is an order-3 homogeneous polynomial function and $P^1$ is defined as before.
\bea\label{Ampthreehalf}
&&P^3\left([\td\bt_1,\td\bt_2],\langle\lmd_1,\lmd_2\rangle\right)\nb\\ &=&a \, \langle\lmd_1,\lmd_2\rangle^3+b \, \langle\lmd_1,\lmd_2\rangle^2 [\td\bt_1,\td\bt_2] \nb\\
&+& c \, \langle\lmd_1,\lmd_2\rangle^1 \, [\td\bt_1,\td\bt_2]^2+ d \, [\td\bt_1,\td\bt_2]^3
\eea
where $a,b,c,d$ are the constant coefficients.

Thus we find that there are altogether six form-factors for the spin 3/2 charged particle. All of the terms can be deduced from Lorentz invariant interactions terms of some lagrangian form. The terms proportional to $P^1$ come from $\bar\psi_{\omega} \, p_3^{\omega}  \, \Sigma_{\mu\nu} \, p_3^\nu \, p_3^\rho\psi_{\rho}  \, A^\mu$ and $\bar\psi^1_{\omega} \, p_3^{\omega}  \, \gamma_\mu  \, p_3^\rho \, \psi_{\rho}  \, A^\mu$.  Two of the terms with coefficients $b,c$ in $P^3$ can be deduced from $\bar\psi_{\mu}  \, \gamma^\kappa  \, \psi^{\mu}  \, A_\kappa$ and $\bar\psi_{\rho}  \,\Sigma^{\mu\nu} k_\nu\, \psi_{\rho} \, A_\mu$. Here, $\Sigma^{\mu\nu}$ is the generator of Lorentz transformations in the vector-spinor Rarita-Schwinger representation.

The remaining two interactions are \\ 
$$\bar\psi^\mu  A_\mu~ p^3_\nu \psi^\nu- \bar\psi^\mu  p^3_\mu ~ A^\nu \psi_\nu$$ 
and 
$$\bar\psi_{a\dot b}p^{\dot b b}_1 p_{b\dot c}^2  A^{\dot c a}~ p^3\cdot \psi- \bar\psi\cdot p^3 ~ A_{a\dot b}p^{\dot b b}_1 p_{b\dot c}^2 \psi^{\dot c a}.$$ 

The there point amplitudes of these  two interactions  are some linear combinations of the terms with $P^3$ factor.  After a simple calculation we get 
\bea
&&(\,2 \langle\lmd_1,\lmd_2\rangle^2   +\, 2\langle\lmd_1,\lmd_2\rangle \, [\td\bt_1,\td\bt_2]+  \, [\td\bt_1,\td\bt_2]^2)\nb\\ &&\times{[\td\bt_1,\td\bt_2]\langle\lmd_2,\lmd_3\rangle \over [\td\bt_2, \td\lmd_3]}\nb 
\eea
and 
\bea
&&(\langle\lmd_1,\lmd_2\rangle^2+ \, 2\langle\lmd_1,\lmd_2\rangle [\td\bt_1,\td\bt_2] + \, 2 \, [\td\bt_1,\td\bt_2]^2) \nb\\
&&\times{\langle\lmd_1,\lmd_2\rangle\langle\lmd_2,\lmd_3\rangle \over [\td\bt_2, \td\lmd_3]}\nb
\eea
respectively. 
All of these interactions are gauge invariant, in the sense that the amplitudes vanish for longitudinal photons, $\epsilon_3^\mu=p_3^\mu$, including the case when the photon is off-shell. If we discard all of the power-counting non-renormalizable terms, i.e. those that have dimension 5 or greater then only the $\bar\psi_{\mu}  \, \gamma^\kappa  \, \psi^{\mu}  \, \epsilon^3_\kappa$ remains, and in spinor form corresponds to a specific choice of coefficients in  (\ref{ampthreehalf}) and (\ref{Ampthreehalf})
\be
\bar\psi_{\mu}  \, \gamma^\kappa  \, \psi^{\mu}  \, \epsilon^3_\kappa = \langle\lmd_1,\lmd_2\rangle [\td\bt_1,\td\bt_2]^2 {\langle\lmd_2,\lmd_3\rangle \over [\td\bt_2,\td\lmd_3]}
\ee

In order to get the amplitudes for other helicity structures for the massive fermions, we do not need to follow the analysis one by one. In fact, form the amplitude with fermions in the lowest weight states of $\mathcal{R}(J^-)$, we can get the amplitudes of all helicity structures by acting it with $\mathcal{R}(J^+)$ to each particle respectively. For example
\bea
&&A\big(({3\over 2},{-3+2n\over 2}),({3\over 2},{-3+2m\over 2}),  (1,-1)\big)\nb\\
&=&(J_1^+)^n (J_2^+)^m A\big(({3\over 2},{-3\over 2}),({3\over 2},{-3\over 2}), (1,-1)\big),\nb\\
\eea
where $n,m = 1,2,3$.

It is possible and interesting from the phenomenological point of view to consider also the interaction of spin 3/2 and spin 1/2 fields with the photon. Some forms of this amplitude were considered by Novaes and Spehler \cite{Novaes} and more recently, in connection with LHC searches for excited quarks by Feng et al \cite{Feng}.

\section{Spinor form for general half spin particles}\label{generalspin}
For the half spin $j$ particles, we can do similar analysis. The fields are in the representation $\mathcal{D}^{(j)}$ of the little group. Then we have
\bea
\mathcal{R}(J^3) u_r&=&\mathcal{D}^{(j)}_{sr}(J^3)u_s\nb\\
\mathcal{R}(J^2) u_r&=&\mathcal{D}^{(j)}_{sr}(J^2)u_s\nb\\
\mathcal{R}(J^1) u_r&=&\mathcal{D}^{(j)}_{sr}(J^1)u_s.
\eea
The wavefunctions in spinor form can also be obtained. First we can get the spinor form of the field for the lowest helicity.  Then for any higher helicity, we can get the wave functions by acting the $\mathcal{R}(J^+)$ operator on the lowest helicity states.

The particle can be obtained by Rarita-Schwinger tensor-spinors $\psi^{\mu_1...\mu_{j-{1\over 2}}}$ \cite{Rarita}, which is symmetry for the vector index and satisfy the Pauli and Fierz \cite{Fierz} constraints $p_{\mu_1} \psi^{\mu_1...\mu_{j-{1\over 2}}}=0$, $\gamma_{\mu_1}\psi^{\mu_1...\mu_{j-{1\over 2}}}=0$. A free field also satisfy the  Dirac equation $(\gamma_\mu p^\mu-m)\psi^{\mu_1...\mu_{j-{1\over 2}}}=0$. Then similar with the spin $3\over 2$ fields,  for vector tensor part, the lowest helicity can only be composed of the $(j-{1\over 2})$-times tensor product by the spin down vectors $\lmd \td\bt$. And the  Dirac equation will constraints the four-components of the  Dirac spinor to be $(\lmd \oplus\td\bt)$ for the positive particles with the spin down helicity. The lowest helicity vector is obtained by symmetry product the tensor part and the  Dirac-spinor part as
\be
u_{-j}=-(\lmd \td\bt)^{j-{1\over 2}} (\lmd \oplus\td\bt).
\ee
The higher helicity states can be obtained as
\be\label{gspinJact}
u_{(-j+n)}=N(j,n) (J^+)^n u_{-j},
\ee
where $n$ in integer and $\in [1, 2j]$ and $N$ is the normalize constant. The manifest form of equations (\ref{gspinJact}) are
\bea
&&u_{-j+n}=N(j,n)\times \nb\\
&&\left(\right.\sum_{i=0}^{[{n\over 2}]}S^n_i\otimes(\lmd \oplus\td\bt)-\sum_{i=0}^{[{n-1\over 2}]}S^{n-1}_i\otimes(\bt \oplus -\td\lmd) \left.\right) .\nb\\
\eea
Here 
$$S^n_i=\mathcal{S}\big((\lmd\td\bt)^{\otimes j-{1\over 2}+i-n}\otimes (\lmd\td\lmd-\bt\td\bt)^{\otimes (n-2i)}(-2\bt\td\lmd)^{\otimes i}\big).$$
$\mathcal{S}$ denote the symmetry tensor product and $[{n\over 2}]$ is the largest integer less than or equal to $n\over 2$. The three point functions of two high spin fermions and one gauge boson can also be obtained directly but with more labor for any helicity structure.

\section{Conclusion}
In this paper, we mainly focus on the massive fields with half-integral spins. We constructed the spinor-form wavefunctions for the free external particles. We also analyze the on-shell amplitudes with two-fermions and one gauge boson. Here the fermions are of the same non-zero mass and the gauge boson is massless. At the beginning, we present the first-order differential form for each generator of the little group. We find this formation play an critical role in obtaining the spinor forms for both free massive fields and amplitudes.

For the half-integral particles in Rarita-Schwinger (RS) framework, we derive the wavefunctions in two steps. First, according to the dirac equation and two constraints in RS framework, we can directly obtain the state of the lowest helicity. Then for the fields of higher helicities, we get their spinor forms by acting the rising operator $\mathcal{R}(J^+)$ on the lowest helicity states.  This procedure are universe for any half-integral spin field. In fact, it can also be  extended to the massive integral spin field with minor modification.

The same procedure is also applicable for the amplitudes. That means we only need to discuss the amplitudes with the lowest helicity structures for the massive fields. For the three-point functions we considered here, according to the linear property for the amplitude with respect to the external fields, in principle we can get the most general forms of the amplitudes which are Lorentz and gauge invariant. Absolutely the constructed amplitudes are the solutions of the seven one-order differential equations which the amplitude satisfy. In this article, we first obtained the spinor form of the on-shell gauge invariant amplitudes for the spin $1\over 2$ field. Each independent  term in the amplitude  corresponds to an interaction. We find the corresponding interactions are also gauge invariant, in the sense that the amplitudes vanish for the case when the external photon is off-shell. We also discuss in detail for the amplitudes for the spin $3\over 2$ particle. By the same methods, we obtain six independent terms in the general amplitude.  Each term can be deduced from a Lorentz invariant interaction. For higher spin fermions, we can also analysis the amplitudes similarly. Further more, all the possible amplitudes with two fermions of  different spin can also be obtained.
\section*{Acknowledgement}

This work is funded by the Priority Academic Program Development of Jiangsu Higher Education Institutions (PAPD), NSFC grant No.~10775067, Research Links Programme of Swedish Research Council under contract No.~348-2008-6049, and the
Chinese Central Government's 985 Project grants for Nanjing University.

We thank Zhang Yun for cross-checking the \\
dimension-5  form-factors  appearing in Section \ref{Spinthreehalf}.
%

\end{document}